\documentclass[runningheads]{llncs}
\usepackage[T1]{fontenc}
\usepackage{graphicx}

\usepackage{caption}

\DeclareCaptionType{listing}[Listing][List of Listings] 

\usepackage{subcaption}

\usepackage{booktabs}

\usepackage{hyperref}

\usepackage{array}

\usepackage{listings}

\usepackage[altpo, epsilon]{backnaur}

\usepackage[usenames,dvipsnames]{xcolor}
\usepackage{mdframed}
\definecolor{LightGray}{gray}{0.9}

\lstdefinestyle{CPP}{
	language=C,
	showstringspaces=false,
	frame=single,  breaklines=true, basicstyle=\scriptsize,
	numbers=left, numberstyle=\tiny, stepnumber=1, numbersep=5pt,%
	backgroundcolor=\color{gray!20},%
	moredelim=[s][\color{MidnightBlue}]{\#}{\
	},
}%

\lstdefinestyle{LISP}{
	language=LISP,
	upquote=true,
	showstringspaces=false,
	frame=single,  breaklines=true, basicstyle=\scriptsize,
	numbers=left, numberstyle=\tiny, stepnumber=1, numbersep=5pt,%
	backgroundcolor=\color{gray!20},%
	moredelim=[s][\color{MidnightBlue}]{'}{\ },
	keywordstyle=\color{black}
}%

\definecolor{delim}{RGB}{20,105,176}
\definecolor{numb}{RGB}{106, 109, 32}
\definecolor{string}{rgb}{0.64,0.08,0.08}

\lstdefinelanguage{JSON}{
	numbers=left,
	numberstyle=\scriptsize,
	frame=single,
	rulecolor=\color{black},
	showspaces=false,
	showtabs=false,
	breaklines=true,
	postbreak=\raisebox{0ex}[0ex][0ex]{\ensuremath{\color{gray}\hookrightarrow\space}},
	breakatwhitespace=true,
	basicstyle=\ttfamily\scriptsize,
	backgroundcolor=\color{gray!20},
	upquote=true,
	morestring=[b]",
	stringstyle=\color{MidnightBlue},
	literate=
	*{0}{{{\color{numb}0}}}{1}
	{1}{{{\color{numb}1}}}{1}
	{2}{{{\color{numb}2}}}{1}
	{3}{{{\color{numb}3}}}{1}
	{4}{{{\color{numb}4}}}{1}
	{5}{{{\color{numb}5}}}{1}
	{6}{{{\color{numb}6}}}{1}
	{7}{{{\color{numb}7}}}{1}
	{8}{{{\color{numb}8}}}{1}
	{9}{{{\color{numb}9}}}{1}
	{\{}{{{\color{delim}{\{}}}}{1}
	{\}}{{{\color{delim}{\}}}}}{1}
	{[}{{{\color{delim}{[}}}}{1}
	{]}{{{\color{delim}{]}}}}{1},
}

\begin{document}
	\counterwithout{lstlisting}{chapter}
\title{Enhancing Performance Monitoring in C/C++ Programs with EDPM: A Domain-Specific Language for Performance Monitoring}
\titlerunning{{Enhancing Performance Monitoring in C and C++ Programs with EDPM}}
%
\author{David Weisskopf Holmqvist \and
Suejb Memeti\orcidID{0000-0003-1608-3181}}
\authorrunning{D. W. Holmqvist and S. Memeti}
%
\institute{Department of Computer Science (DIDA), \\
	Blekinge Institute of Technology, Karlskrona 371 79, Sweden \\
\email{daae19@student.bth.se} and   
\email{suejb.memeti@bth.se}\\}
\maketitle              
\begin{abstract}
The utilization of performance monitoring probes is a valuable tool for programmers to gather performance data. However, the manual insertion of these probes can result in an increase in code size, code obfuscation, and an added burden of learning different APIs associated with performance monitoring tools. To mitigate these issues, EDPM, an embedded domain-specific language, was developed to provide a higher level of abstraction for annotating regions of code that require instrumentation in C and C++ programs. This paper presents the design and implementation of EDPM and compares it to the well-known tool PAPI, in terms of required lines of code, flexibility in configuring regions, and performance overhead. The results of this study demonstrate that EDPM is a low-resolution profiling tool that offers a reduction in required lines of code and enables programmers to express various configurations of regions. Furthermore, the design of EDPM is such that its pragmas are ignored by the standard compiler, allowing for seamless integration into existing software processes without disrupting build systems or increasing the size of the executable. Additionally, the design of the EDPM pre-compiler allows for the extension of available performance counters while maintaining a high level of abstraction for programmers. Therefore, EDPM offers a promising solution to simplify and optimize performance monitoring in C and C++ programs.

\keywords{performance monitoring \and domain-specific languages \and language abstractions \and compilers}
\end{abstract}
%
%

\section{Introduction}
\label{sec:introduction}

Software optimization is a crucial process in the development of software systems, which aims to produce a more efficient and resource-friendly output. The effectiveness of optimization can only be ensured through empirical measurements of performance. Performance counters play a vital role in enabling users to extract essential information about the performance behavior of their programs, and to empirically verify the impact of code changes.

However, the process of collecting such data has been challenging and prone to errors for application developers. The accessibility of hardware performance counters varies depending on the hardware manufacturer, and obtaining access is often not straightforward. Fortunately, several tools have been developed to unify access to such counters, mitigating this challenge for developers.

PAPI (Performance API) \cite{papi:2000} is a widely used tool that facilitates the manual insertion of performance monitoring probes into the source code. Similarly, the Performance Counters Library (PCL) \cite{pcl:1998} offers similar functionality but with some limitations. While manual tools like PAPI and PCL provide programmers with the flexibility to collect performance counters selectively and precisely in different regions of the program, they also introduce foreign code that can obfuscate the application logic.

Alternatively, automated tools like gprof\cite{gprof:1982}, valgrind\cite{valgrind:2003}, and TAU\cite{shende2006tau} do not require manual insertion of performance monitoring probes into the source code, freeing programmers from the burden of writing code related to collecting performance counters. However, automated tools often provide a uniform collection of counters throughout the whole program execution, making it challenging to identify discrete performance concerns in specific program regions.

This paper presents the design and implementation of an Embedded Domain-specific language for Performance Monitoring (EDPM) of C and C++ programs. The purpose of EDPM is to simplify the process of performance monitoring in software development by enabling programmers to annotate regions of code for monitoring using a high-level language, thereby reducing programming effort. Unlike existing alternatives, EDPM offers the added advantage of permitting nested code region hierarchies with varying performance counters.

EDPM is designed to allow for easy extension, such that additional performance counters can be collected by incorporating multiple back-ends. The prototype presented in this paper demonstrates that EDPM strikes a balance between flexibility and ease-of-use, thereby mitigating the typical increase in programming effort and line-of-code that is associated with manual profiling.

The main contributions of this paper include:
\begin{itemize}
	\item The design and implementation of an Embedded Domain-specific Language for Performance Monitoring, incorporating pragma-based compiler directives and facilitating seamless integration with C/C++ programming languages.
	\item The creation of a pre-compiler capable of transforming the high-level compiler directives into low-level source code appropriate for the target system.
	\item 	A comparative analysis of the proposed solution with existing manual profiling tools, leveraging the current state of the art.
\end{itemize}

The following is the organizational structure of this paper: Section \ref{sec:background-rw} provides a review of relevant background information and offers a comprehensive summary and comparative analysis of related works; Section \ref{sec:edpm} introduces the Embedded Domain-specific Language for Performance Monitoring (EDPM); Section \ref{sec:evaluation} presents the evaluation of EDPM, providing empirical evidence of its effectiveness; and lastly, Section \ref{sec:conclusion_fw} concludes this paper and outlines potential directions for future research.

\section{Background and related work}
\label{sec:background-rw}

This section presents relevant background information about performance monitoring and domain-specific languages to provide a context for the later sections. Furthermore, it summarizes and comparatively analyses the related works.

\subsection{Performance Monitoring}
\label{sec:perfmon}

Performance monitoring is a crucial activity that involves generating performance measurements to describe the behavior of a software system. Various approaches are available to programmers, each with its own set of advantages and disadvantages. For instance, automated tools such as gprof \cite{gprof:1982} and valgrind\cite{valgrind:2003} facilitate the collection of resource usage information and other relevant properties during program execution. However, such instrumentation tools generally consider the entire program and collect the same set of metrics throughout execution.

In contrast, tools like PAPI\cite{papi:2000}, PCL\cite{pcl:1998}, and LIKWID\cite{likwid2012} provide developers with more granular control over the code regions and specific metrics to be measured. However, this control comes at the cost of convenience, as application developers must include performance monitoring code within their application code. For example, Listing \ref{listing:collecting-data-and-instruction-misses-papi-ll} provides an example of counting data and instruction cache misses at various call sites of \texttt{transpose} and \texttt{matmul} functions (highlighted lines), showcasing the many possibilities for configuring regions of interest and selecting counters.

\captionof{listing}{Collecting data and instruction cache misses with PAPI}
\label{listing:collecting-data-and-instruction-misses-papi-ll}
\includegraphics[width=\linewidth]{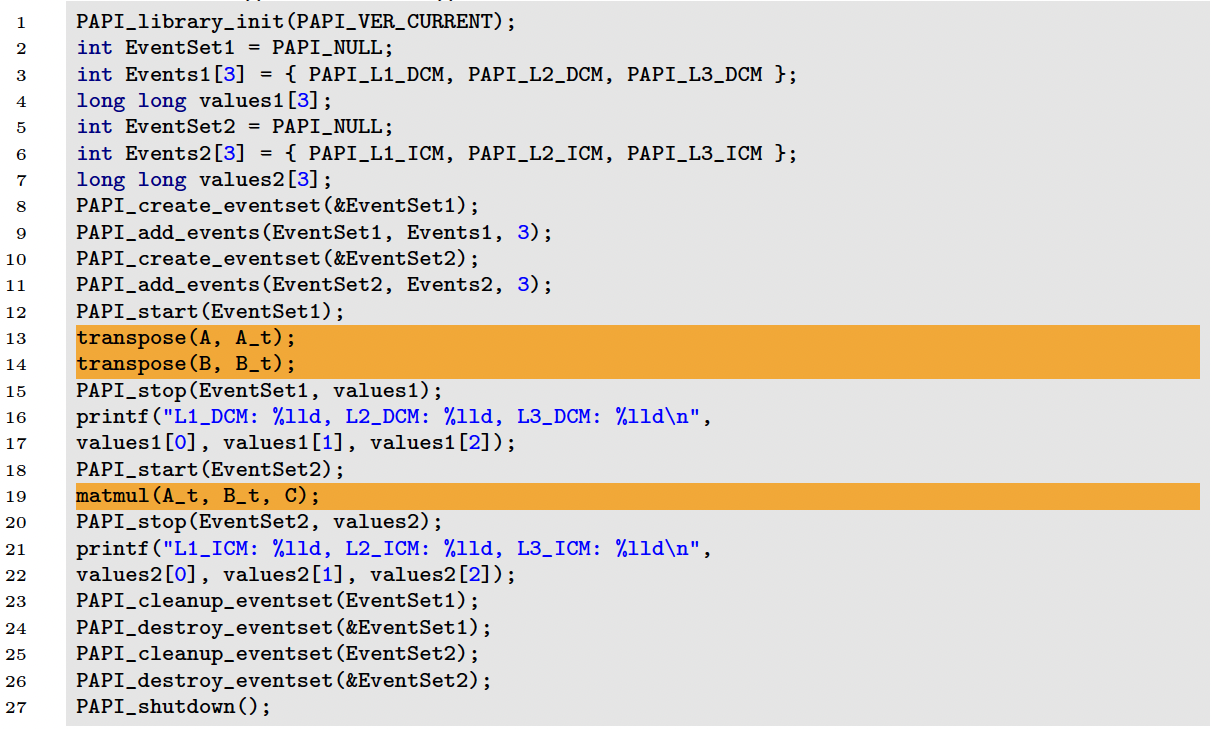}


What is apparent from Listing \ref{listing:collecting-data-and-instruction-misses-papi-ll} is the fact that incorporating performance monitoring code (lines 1-12, 15-18, and 20-27) can overwhelm the application logic (lines 13, 14, and 19).  Utilizing tools that offer more granular control requires proficiency in their usage and API, which presents an additional hurdle.

The code above employs the low-level API, however, PAPI offers a more user-friendly high-level API\cite{papi:2020}, which aligns with the concept of defining regions of interest, as demonstrated in Listing \ref{listing:collecting-with-papi-hl}. The list of counters of interest is specified through a \texttt{PAPI\_EVENTS} environment variable. While this approach allows for modifying the counters without recompiling the application, it also poses the problem of collecting the same set of counters for all designated regions during program execution. The trade-off between convenience and control is evident in this example; with the increase of the abstraction level, programmers are relieved of the responsibility of initializing libraries and variables.

\captionof{listing}{Collecting counters with PAPI's high-level API}
\label{listing:collecting-with-papi-hl}
\includegraphics[width=\linewidth]{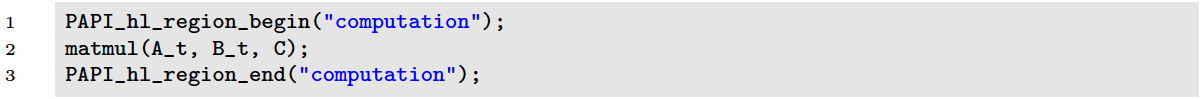}


The Performance Counter Library\cite{pcl:1998} is similar to PAPI, and also exposes a low-level interface for starting, stopping and reading from counters collected in regions. It supports nested regions with the limitation of any single nesting level being unable to accommodate for dynamic sets of counters.

Opari \cite{mohr2001towards} is a performance monitoring tool, which utilizes the POMP monitoring library for measuring different performance related counters. Similarly to EDPM, Opari allows developers to annotate specific regions of OpenMP, for which to collect performance counters. In comparison to EDPM, the Opari API is not based on pragmas even though it targets OpenMP applications. The ompP \cite{furlinger2008ompp} is built on top of Opari to abstract away some of the details and make the profiling process easier for developers. Similarly to EDPM, in ompP there is an \textit{enter} and \textit{exit} API function that wraps a region of interest. 

\subsection{Domain-specific Languages}
\label{sec:dsl}

Abstraction is often deemed the most critical aspect of good software\cite{handbook-dsl:1997}. This notion is evidenced by the myriad of software design concepts that lack a corresponding hardware entity yet serve to reduce software engineering complexity. Examples of these concepts include classes, modules, and higher-order functions. Such concepts exist in general-purpose programming languages (GPLs), enabling programmers to create applications for diverse domains. However, a distinct class of languages, known as domain-specific programming languages (DSLs), employ domain-specific notations to succinctly express entities and their functions.

In the domain of performance analysis and optimization, examples of DSLs include  the TigrisDSL \cite{tigris:2021}, which allows programmers to specify events to be monitored by writing user definitions of so-called relevance criteria, which are used in its first phase to collect coarse-grained metrics. The collected data and the user-defined relevance specifications are then used in a second phase to identify relevant parts of the software system to be monitored for more fine-grained metrics. This two-phase approach decreases the performance monitoring overhead, which is crucial for runtime adaptation of a software system.

Another example is OpenMP\cite{openmp:1998} in the domain of parallelization of software. OpenMP is a set of compiler directives and runtime procedures that allow programmers to express shared-memory parallelism. OpenMP promised portability to allow application developers to adopt the shared-memory programming model; in the process, it is clear that a DSL was developed with constructs to express the different actions in the domain of data- and task-parallelism.

Performance monitoring is vital in High Performance Computing (HPC), where heterogeneous architectures pose significant challenges for developer. \\ANTAREX \cite{antarex:2019} offers a novel approach to performance optimization by separating developers from the optimization process and entrusting support staff at HPC centers to handle it. The project has developed an aspect-oriented DSL that automates parallelization, offloading, and performance analysis and tuning. Domain-level experts can express functional properties like energy efficiency and performance using the ANTAREX DSL, enforced through runtime code generation using a "collect-analyze-decide-act" loop. 

 HSTREAM \cite{hstream:2018} introduces a compiler extension based on pragma directives that enables programmers to annotate parallel regions of code, in combination with a runtime system that, with the help of the code that is generated from the annotations, distributes the workload across the different processing units available on the architecture. In the paper, the authors show that HSTREAM, when compared to code that solely uses CPUs or GPUs, both provides greater performance and lower programming complexity.

EDPM provides a user-friendly interface for collecting performance counters without increasing programming complexity or disrupting application logic. Unlike other domain-specific languages such as ANTAREX, TigrisDSL, and HSTREAM, which aim to achieve different goals, EDPM employs linguistic abstraction to reduce programming complexity for performance monitoring. EDPM's use of pragma directives for generating code is similar to OpenMP and HSTREAM. Although currently relying solely on PAPI as its backend, EDPM has the potential to generate code for other backends, as discussed in Section \ref{sec:conclusion_fw}. EDPM offers a valuable tool for programmers seeking to optimize program performance.

\section{EDPM: C/C++ Language Extension for Ergonomic and Flexible Performance Monitoring}
\label{sec:edpm}

EDPM uses pragma directives to mark specific regions of code and specify which counters to collect for each region. This technique is similar to that used by OpenMP and HSTREAM, and has the advantage of not interfering with regular program compilation. The user must still annotate points of initialization and termination, as demonstrated in listing \ref{listing:edpm-annotated}.

EDPM precompiles annotated source code and generates code for a specific performance monitoring backend (e.g., PAPI). EDPM offers a high degree of flexibility by allowing programmers to specify different counters for different regions in the same program. EDPM supports lexical nesting of regions (a feature inspired by the TAU \cite{shende2006tau} performance monitoring tool), enabling programmers to specify properly nested (where a region is fully contained within an outer region) or overlapping (where only one of the regions' boundaries lies within another region) regions.  This flexibility makes it possible to handle diverse scenarios, such as an application that predominantly uses memory in one part and CPU in another. However, EDPM currently does not support dynamic nesting, which refers to nesting that becomes apparent during program execution.

\captionof{listing}{EDPM-annotated region that collects counters related to memory}
\label{listing:edpm-annotated}
\includegraphics[width=\linewidth]{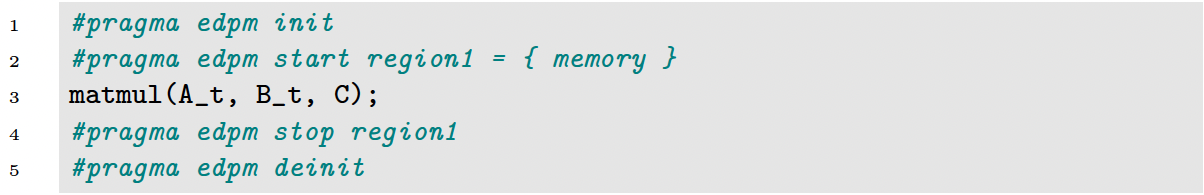}

EDPM provides several directives for marking code regions and collecting counters. The directives are lines prefixed with \texttt{\#pragma edpm} and include \texttt{init/ deinit} (to initialize and deinitialize the counter collection mechanism), and \texttt{start/stop} (to start and stop a region to be monitored). The \textit{start} directive takes a comma-separated list of region descriptions consisting of clauses specifying the counters to collect. The available counters are grouped by type, as shown in table \ref{table:counter-table}, and correspond to a selection of PAPI's API counters \cite{papi:2000}. 

\captionof{listing}{Excerpts of a matrix multiplication application that uses EDPM}
\label{listing:matmul}
\includegraphics[width=\linewidth]{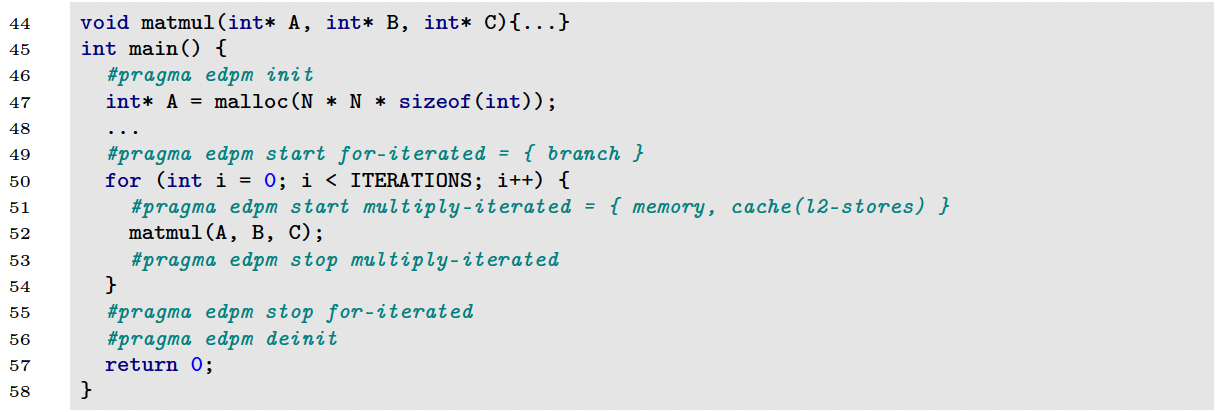}

\begin{table}[!htb]
	\centering
	\begin{tabular}{ m{2cm} m{31em} }
		\toprule 
		\textbf{Type} & \textbf{Counters} \\
		\midrule
		cpu & cycles, instructions \\
		memory & loads, stores \\
		floating-point & instructions, operations, multiply, add, divide, sqrt, inverse \\
		vector & single-precision, double-precision \\
		branch & unconditional, conditional, taken, not-taken, mispredicted, correctly-predicted \\
		cache & invalidation, l1-data, l2-data, l3-data, l1-instructions, l2-instructions, l3-instructions, l1-loads, l2-loads, l1-stores, l2-stores \\
		\bottomrule
	\end{tabular}
	\caption{List of counters available for collection, grouped by type}
	\label{table:counter-table}
\end{table}

A concrete example is described here to display how EDPM could be used to collect counters. We assume the presence of a file \texttt{matmul.c} as shown in listing \ref{listing:matmul}. In this file there are two regions, named in a describing manner to communicate the purpose of the regions: \textit{for-iterated} that collects counters related to branching around the for loop, and \textit{multiply-iterated} that collects counters related to memory and specifically L2 cache stores.

Listing \ref{listing:json-output} presents an example of the JSON output file produced by EDPM. It is noteworthy that the entries in the output file correspond to the names of the regions of interest, as exemplified by the name \textit{multiply-iterated} in line 4, which corresponds to the region inside the loop body shown in lines 8-10 of Listing \ref{listing:matmul}. Since regions may be executed multiple times, as in the case of the \textit{multiply-iterated} region, which is executed inside a loop, a \textit{temporal-id} field is included in the JSON file to distinguish between runs. For instance, the value \textit{"temporal-id" : 0} in line 6 indicates that the collected counters correspond to the first iteration of the loop. Users can utilize their preferred JSON-aware tools to filter out specific regions and sum the counters to obtain the desired total.


\captionof{listing}{Excerpt of the JSON file output when running EDPM with \texttt{matmul.c}}
\label{listing:json-output}
\includegraphics[width=\linewidth]{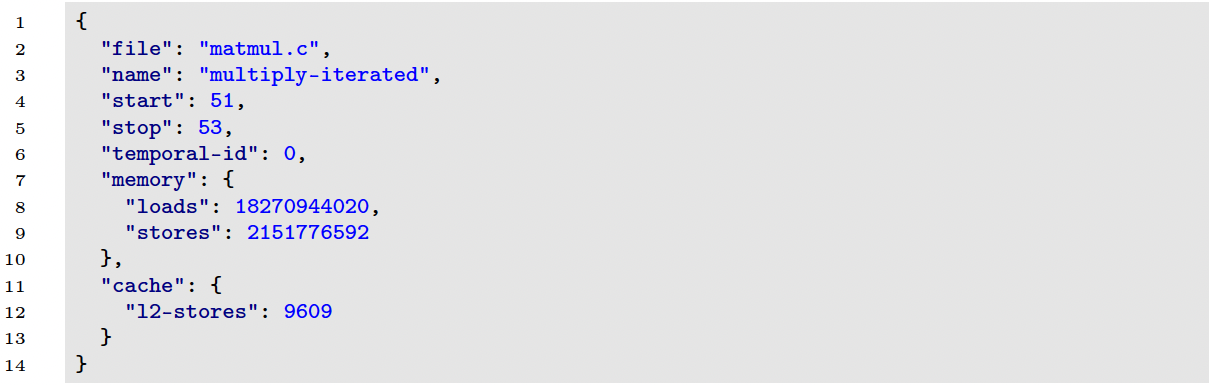}

\subsection{Design \& Implementation}
\label{sec:design-implementation}

EDPM is a precompiler for C/C++ programs that follows the standard phases of a compiler. These phases include the reader, which combines the lexer and parser, semantic analysis, the intermediate representation, and code generation. The subsequent sections provide a brief overview of each of these phases.

\subsubsection{Reader}
\label{sec:reader}

The reader, consisting of the lexer and parser, receives a C/C++ source code file and produces a list of directive structures. These structures include the directive's type (init, deinit, start, or stop), name (applicable only for start and stop directives and refers to the region), a collection of clauses, and the line number from the source code file where it was obtained. A clause is also a structure type that comprises its type and counters, which align with Type and Counters columns in Table  \ref{table:counter-table}. EDPM only considers lines prefixed with \texttt{\#pragma edpm} and currently is not aware of the syntax or semantics of C/C++ programs.

The EDPM formal grammar, is intentionally designed to be simple, which is a key advantage of the language. Enhancing EDPM with a static analysis step, could further simply the grammar by eliminating the \textit{init} and \textit{deinit} pragmas.

\subsubsection{Semantic Analysis and Intermediate Representation}
\label{sec:semantic-analysis}

The semantic analysis generates a list of \texttt{ir-directive} structures, a list of \texttt{block} structures, and a region-table that maps region names to \texttt{region-info} structures. The list of directives undergoes several small passes to ensure semantic correctness. These passes check that only one init and deinit directive is provided, unique region names have been chosen for each region, and that unique types and counters have been specified for each start directive.

Following these passes, a small pass called "normalize" is executed to ensure that all types and counters provided by the user are recognized by EDPM, as per table \ref{table:counter-table}. This pass also expands any empty counter specifications, which results in a list of expanded directives that EDPM recognizes.

The list of expanded directives is then passed to a stack-based algorithm called "collect-blocks" that identifies all consecutive blocks by pushing counters onto the stack when encountering a start directive and popping them when encountering a stop directive. When the last set of counters is popped from the stack, a \texttt{block} structure type is created with the set union of all counters found. This enables users to collect the same set of counters in different regions within the same block. Information about the start and stop positions of the block, as well as reserved variable names for \textit{eventsets} (an integer) and counters (a static array of \textit{long long}), are also collected.

Next, the expanded directives are processed once more with "collect-regions," another stack-based algorithm that assigns indices into the values array variable. This algorithm uses the information collected in the previous pass and results in a region-table that maps region names to \texttt{region-info} structures, and a list of \texttt{ir-directive} structures.

The \texttt{region-info} structure contains the start and stop positions of the region, a \texttt{block-index} structure, another reserved variable name for counters (a static array of \textit{long long}), and a reserved variable name for a temporal identifier that lets programmers differentiate between different executions of the same regions. The \texttt{block-index} structure contains a means of accessing the corresponding \texttt{block} structure and the assigned indices. While a block must allocate space for all the counters it needs to count, a region allocates space only for the counters it requires, consulting the block's values using the assigned indices.

The \texttt{ir-directive} is the intermediate representation for EDPM, containing information about the entity referenced by the directive, its type, an identifier reference, and a position representing the line number in the source code file where generated code will reside. When encountering a start directive during evaluation of \texttt{collect-regions} with an empty stack, a block is started (see Listing \ref{listing:start-intermediate-pause}, lines 1-3). However, when the stack is not empty, the current counter values maintained by the block must be updated/accumulated, and the collection of counters is paused to prevent EDPM actions from influencing the counters and skewing performance monitoring results. The current counter values are then copied over into the region's array as a starting value, and finally, the block is resumed again (see Listing \ref{listing:start-intermediate-pause} lines 4-6).

\captionof{listing}{Intermediate directives generated upon encountering a start directive when the maintained stack is empty (lines 1-3) and non-empty (lines 1-6)}
\label{listing:start-intermediate-pause}
\includegraphics[width=\linewidth]{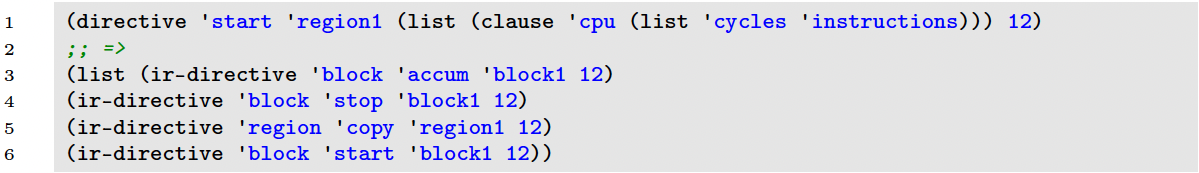}

\subsubsection{Code Generation}
\label{sec:codegen}

EDPM uses information from the semantic analysis phase to generate object, header, and source code files. The bindings collected during semantic analysis, such as reserved variable names for event sets and counters, are used to create these files. Additionally, auxiliary bindings like global integers and file pointers are generated. EDPM's code generation phase generates code specific to the backend (in this case, PAPI) by utilizing the low-level API for PAPI. The library is initialized at the \texttt{"\#pragma edpm init"} point, where event sets are created and counters are added to those event sets. At the \texttt{"\#pragma edpm deinit"} point, the event sets are cleaned up and destroyed, and the library is shutdown. The resulting output is a file-spec structure for each file type, which is later passed to the runner for compilation and counter collection.

\section{Empirical evaluation}
\label{sec:evaluation}

In this section, we evaluate EDPM in terms of code lines and performance overhead. We use a program with three functions, each containing four code blocks for performance analysis. We assess four configurations: E1 (which involve regions around function calls), E2 (regions around code blocks), E3 (properly nested regions around the code blocks), and E4 (alternating overlapping regions around the code blocks). The monitored code blocks involve naive matrix multiplication on 512x512 matrices. For each configuration, we compare EDPM and PAPI's low-level and high-level APIs in terms of programmability (lines of code and region configurations) and performance overhead.



\begin{figure}
	\centering
	\begin{subfigure}[t]{.5\textwidth}
		\centering
		\includegraphics[width=\linewidth]{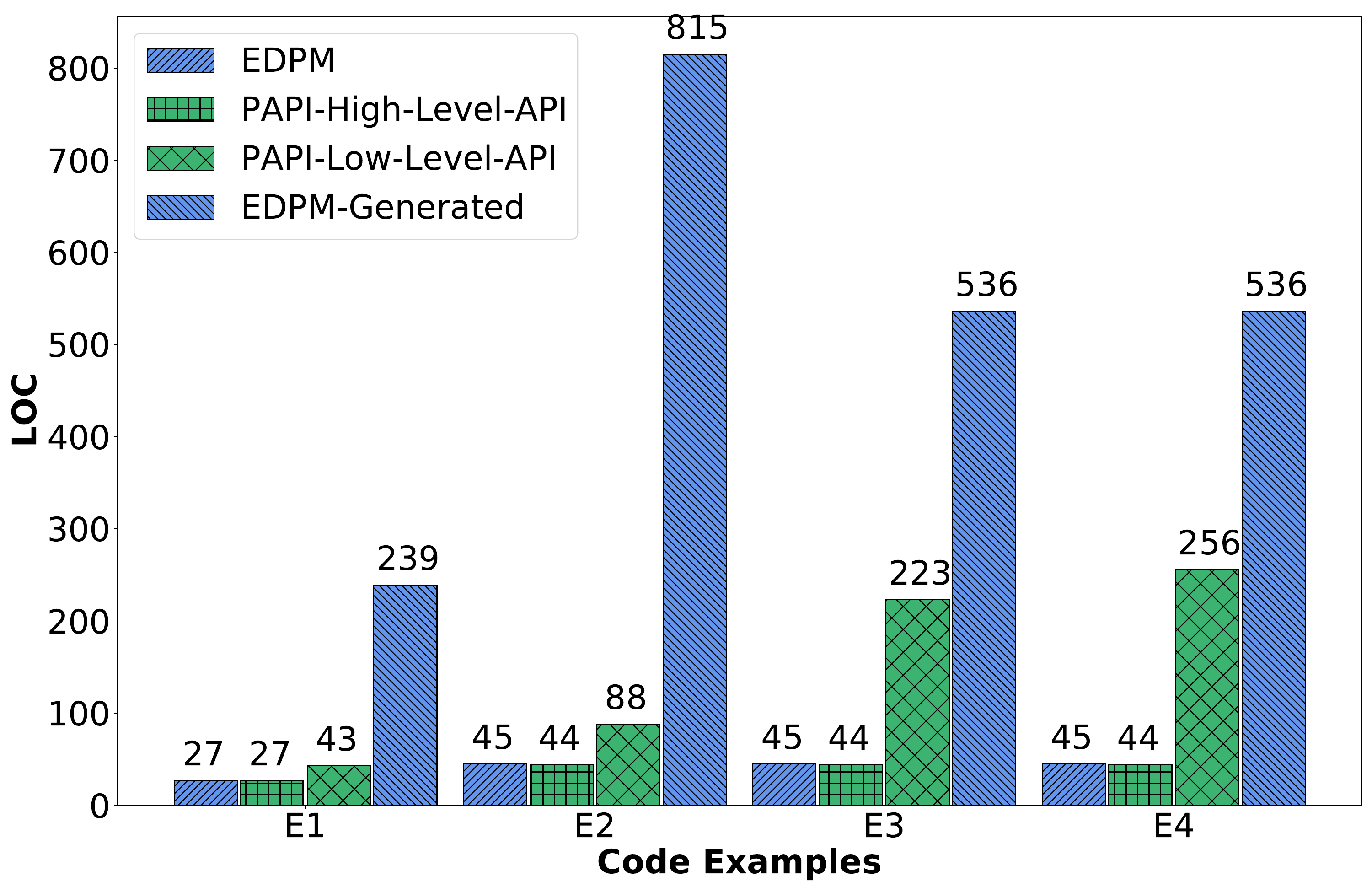}
	\end{subfigure}%
	\begin{subfigure}[t]{.5\textwidth}
		\centering
		\includegraphics[width=\linewidth]{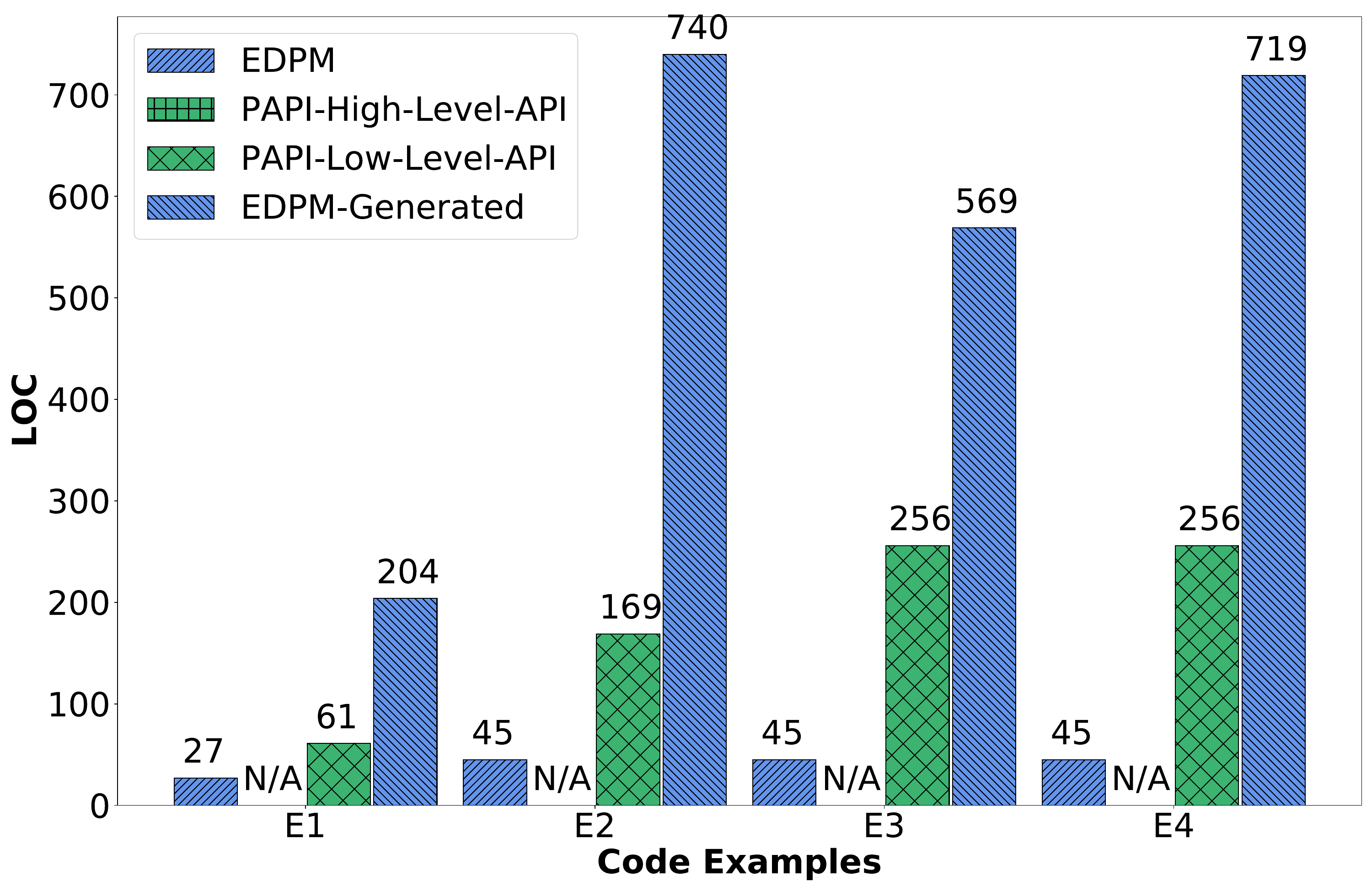}
	\end{subfigure}
	\caption{Comparison of EDPM and PAPI for the static (left) and dynamic (right) code examples with respect to lines of code.}
	\label{fig:loc}
\end{figure}

Figure \ref{fig:loc} depicts the comparison of LOC required for the EDPM, PAPI high- and low-level API, and EDPM-generated code for the Static and Dynamic sample sets. Notably, EDPM demonstrates consistent behavior, requiring the same LOC for both sample sets. Conversely, PAPI high-level API fails to provide any data for the examples in the dynamic set, indicating its inadequacy for such scenarios, whereas for the static set it results with similar lines of code as EDPM. In contrast, the PAPI low-level API requires more effort and significantly more LOC compared to EDPM in both the Static and Dynamic sample sets.

Furthermore, on average, the EDPM-generated code contains approximately 12.69 and 13.15 times more LOC than the original EDPM code in the Static and Dynamic sample sets, respectively. The increase of LOC for the EDPM-generated code compared to the PAPI low-level APi is mainly because EDPM has to generalize and be flexible for future added features.

We may conclude that the LOC for EDPM and the high-level API provided by PAPI for the Static sample set are very similar. However, for the Dynamic sample set, since PAPI's high-level API does not permit having differing sets of counters for different regions in the same program execution, EDPM can be compared to the low-level API of PAPI, which allows this particular expression of configuration. The LOC required for the low-level API is approximately 3.3 to 4.3 times that of the equivalent using EDPM. 

\begin{figure}
	\centering
	\begin{subfigure}[t]{.5\textwidth}
		\centering
		\includegraphics[width=\linewidth]{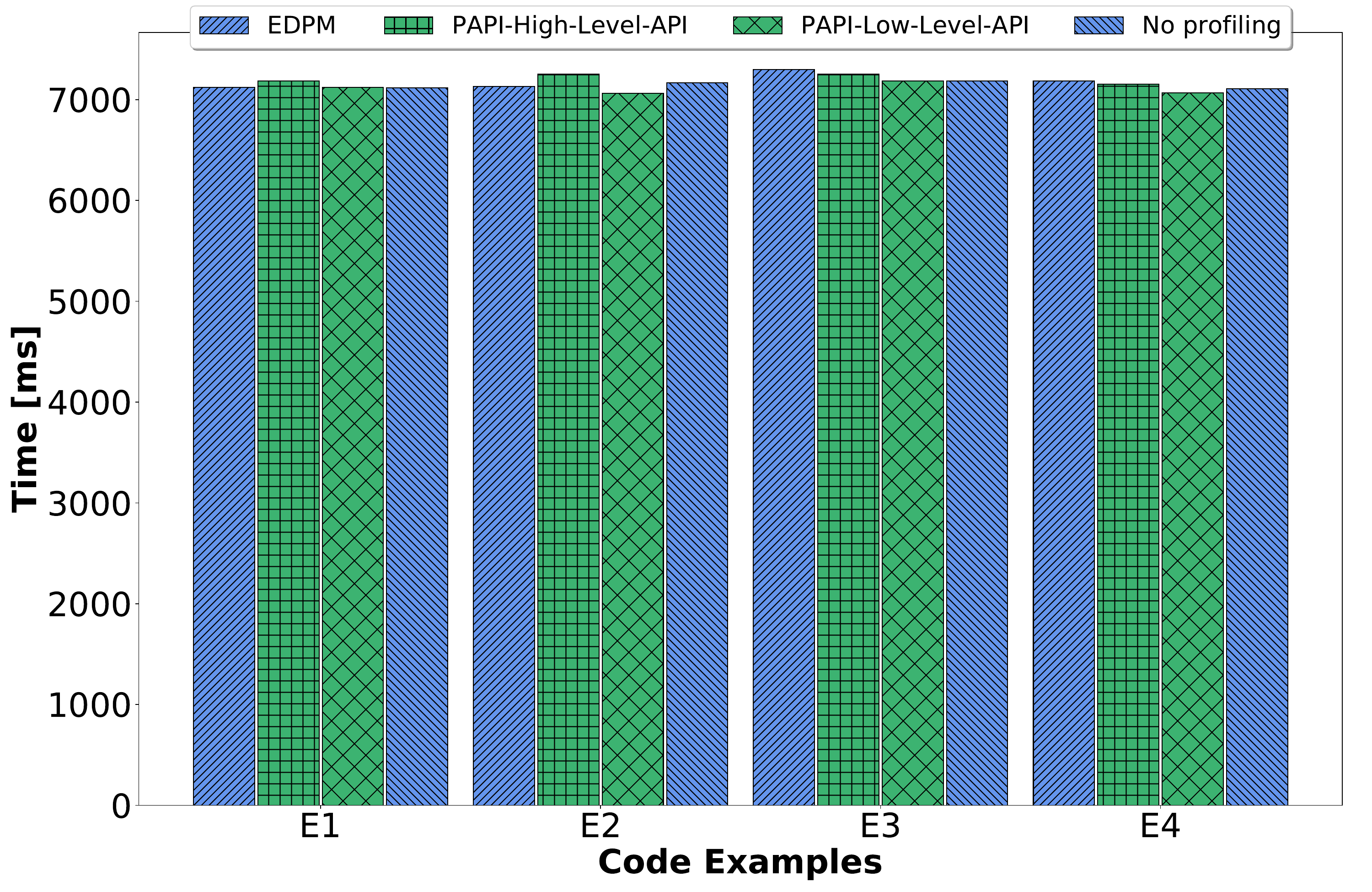}
	\end{subfigure}%
	\begin{subfigure}[t]{.5\textwidth}
		\centering
		\includegraphics[width=\linewidth]{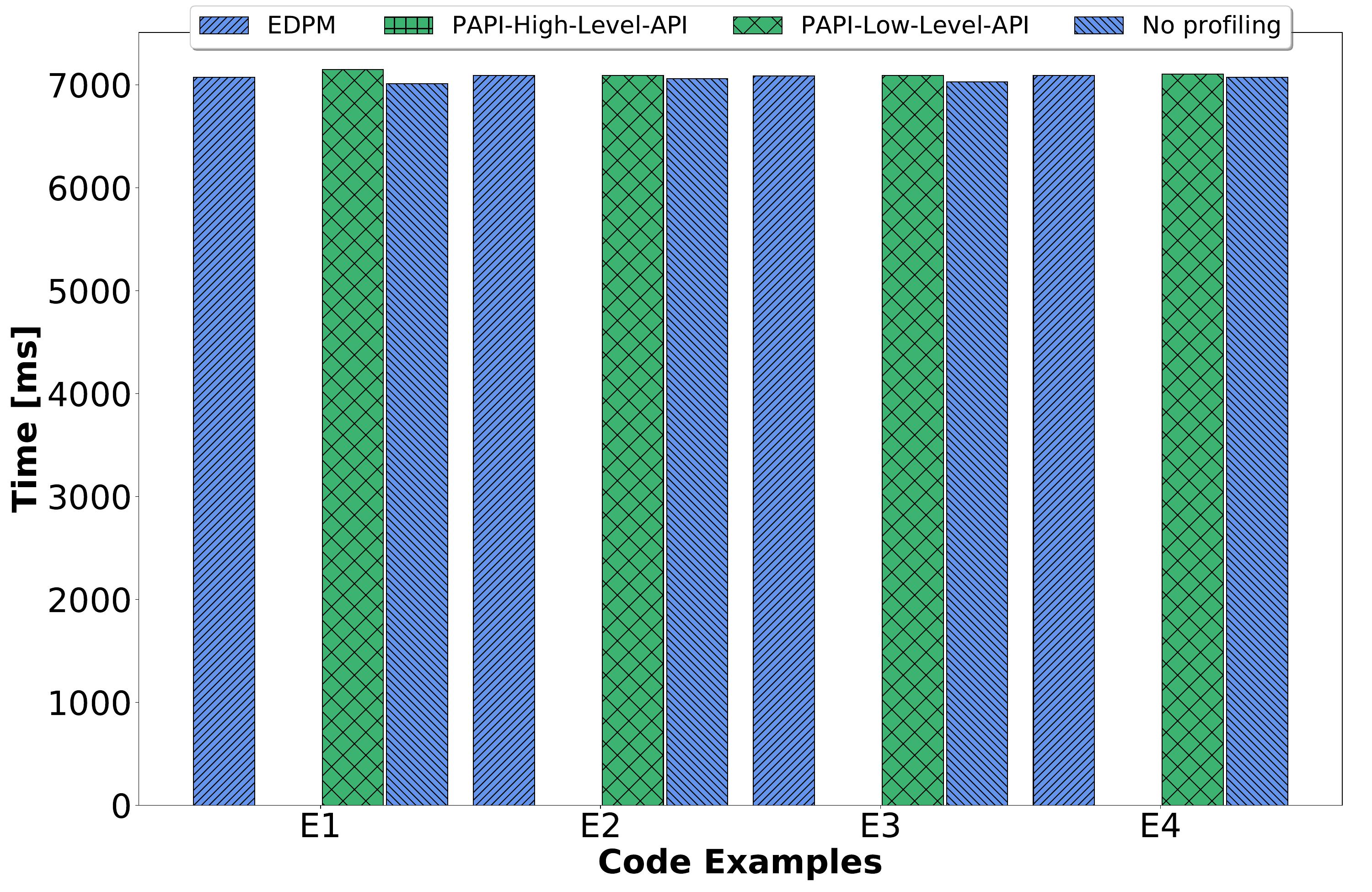}
	\end{subfigure}
	\caption{Comparison of EDPM and PAPI for the static (left) and dynamic (right) code examples with respect to execution time.}
	\label{fig:ms}
\end{figure}

Figure \ref{fig:ms} depicts the execution time of EDPM, PAPI high-level API, and PAPI low-level API compared to the original code that incurs no profiling. Each configuration was executed 10 times, and the results report the average values. One may observe from the figure that no significant performance overheads are added when profiling is enabled, which makes EDPM low-resolution profiling tool, a characteristic that is inherited by the PAPI library.

As described in Section \ref{sec:edpm}, EDPM allows for the addition of new backends by creating a mapping from the intermediate representation to equivalent source code using the new backend, which enables potential source-code level performance monitoring tools to target programmers already familiar with EDPM by implementing necessary translations. If new backends require hooking into different phases of the precompiler, EDPM would have to evolve to provide the necessary flexibility. For programmers using EDPM, the level of abstraction remains constant regardless of changes to the backends or precompiler. 

EDPM holds an advantage over PAPI in that when standard C/C++ compilers process source code annotated with EDPM-specific pragmas, these pragmas are ignored, leading to no increase in the executable file size. In contrast, when using PAPI the size of the target executable file increases. Integrating comparable features using PAPI or other manual profiling tools would require adapting the application logic, which would incur additional time and effort that is not reflected in the presented results related to LOC.

\section{Conclusion and future work}
\label{sec:conclusion_fw}

The authors conclude that EDPM offers comparable user-friendliness to PAPI's high-level API for collecting a static set of counters, while also providing the ability to collect different sets of counters in various regions, which the high-level PAPI API cannot do. Compared to PAPI's low-level API, EDPM offers benefits such as reduced LOC and ease-of-use similar to PAPI's high-level API, while also providing the flexibility and possibilities of the low-level API.

Additionally, EDPM offers high-level abstractions for various configurations for performance monitoring regions. It can be integrated with existing source code without increasing the maintenance costs or disrupting the current build system. To collect and analyze performance data for a new software iteration, the source code can be passed to EDPM without modification.

Moreover, EDPM's design, which includes backends and an intermediate representation, allows for the implementation of new backends to enhance its feature set without reducing the level of abstraction for the end-user.

Future work can explore several areas. One possibility is to support dynamic nesting, as discussed in Section \ref{sec:edpm}. Furthermore, EDPM could be enhanced with the functionality to monitor multi-threaded applications.

Currently, PAPI serves as EDPM's only backend. To broaden counter options and platform availability, new backends could be integrated. If these introduce unique requirements, such as specific precompiler phases, implementation mechanisms would be needed. EDPM could also be extended to include a generic backend-definition framework, acting as a plugin to facilitate the creation of new monitoring tools.

While energy efficiency in hardware has received considerable attention in recent years, less attention has been directed towards energy efficiency in software\cite{energy-efficiency:2012}. PAPI can collect system-wide energy usage\cite{energy-papi:2012}, and EDPM could easily be extended to support this feature. 

Finally, instead of focusing on program-local counters, we could examine system counters to assess the load that specific regions place on the system.

\bibliographystyle{splncs04}

\end{document}